# Lawful but Awful: Evolving Legislative Responses to Address Online Misinformation, Disinformation, and Mal-Information in the Age of Generative AI

Simon Chesterman*

*"Fake news" is an old problem. In recent years, however, increasing usage of social media as a source of information, the spread of unverified medical advice during the Covid-19 pandemic, and the rise of generative artificial intelligence have seen a rush of legislative proposals seeking to minimize or mitigate the impact of false information spread online. Drawing on a novel dataset of statutes and other instruments, this article analyses changing perceptions about the potential harms caused by misinformation, disinformation, and "mal-information". The turn to legislation began in countries that were less free, in terms of civil liberties, and poorer, as measured by GDP per capita. Internet penetration does not seem to have been a driving factor. The focus of such laws is most frequently on national security

* David Marshall Professor of Law at the National University of Singapore and Senior Director of AI Governance at AI Singapore. I'm grateful to Jiang Yu Hang, Shireen Lee, Elizabeth Ong, Clarie Sng, and Sripratak Thanakorn for invaluable research assistance. Thanks also to Fakhar Abbas, Chen Dawei, Stuart Derbyshire, Hahn Jungpil, Tristan Koh, Markus Labude, Eka Nugraha Putra, Amelie Roediger, Soh Kai Xin, Araz Taeihagh, Jun Yu, and Audrey Yue for comments on earlier versions of the text. This research is supported by the Ministry of Education, Singapore, under its MOE AcRF TIER 3 Grant (MOE-MOET32022-0001). The author is not aware of any conflicts of interest in connection with this article. Misinformation and disinformation in the published version are the responsibility of the author alone.

*broadly construed, though 2020 saw a spike in laws addressing public health. Unsurprisingly, governments with fewer legal constraints on government action have generally adopted more robust positions in dealing with false information. Despite early reservations, however, growth in such laws is now steepest in Western states. Though there are diverse views on the appropriate response to false information online, the need for legislation of some kind appears now to be global. The question is no longer whether to regulate "lawful but awful" speech online, but how.*

*Keywords: misinformation, disinformation, free speech, censorship, deepfake, generative artificial intelligence*

For as long as humans have communicated, we have lied to one another. Though former US President Donald Trump sometimes claims to have coined the term,[1] 'fake news' was a headline in the *New York Times* more than a century ago.[2] The history of propaganda is far older. An early example was Egypt's Ramesses the Great, who decorated temples with monuments to his tremendous victory in the 13th century BCE Battle of Kadesh. (The outcome was, at best, a stalemate.)[3] Efforts to limit the harms associated with false information are almost as old, counterbalanced by the emergence of free speech as a political cause and, more recently, as a human right.[4] The digital age has amplified these tensions, with the ability to communicate — for good or ill — on a scale and at a speed previously unimaginable.[5]

This article considers the emergence and evolution of laws intended to reduce the impact of false information shared online, focusing on a novel dataset of 151 national laws adopted since 1995. The number of such laws more than tripled between 2016 and 2023. In 2016,

---

[1] Callum Borchers, *Trump Falsely Claims (Again) that He Coined the Term "Fake News"*, WASH. POST, 26 October 2017.

[2] Capt. Bartlett, *"Fake" News for Spain*, N.Y. TIMES, 6 October 1901.

[3] Boyo G. Ockinga, *On the Interpretation of the Kadesh Record*, 62(123-124) CHRONIQUE D'EGYPTE 38 (1987).

[4] *See* generally ERIC BARENDT, FREEDOM OF SPEECH (2007).

[5] JACOB MCHANGAMA, FREE SPEECH: A HISTORY FROM SOCRATES TO SOCIAL MEDIA (2022).

less than one-fifth of countries had such legislation on the books; by 2023, it was more than half.[6]

Various factors appear to have influenced this turn to law. In addition to the general increase in online communications relative to traditional mass media such as newspapers and television, the 2016 election of President Trump and Britain's referendum in the same year to leave the European Union (Brexit) were linked to allegations of manipulation of users on social media by Cambridge Analytica and others.[7] Secondly, the Covid-19 pandemic saw dozens of countries adopt laws related to public health misinformation in particular beginning in 2020.[8] Thirdly, the rise of generative AI, especially after the release of ChatGPT in November 2022, raised the spectre of ever-greater quantities of ever more realistic fake information — concerns exacerbated by the coincidence of 2024 being the first time in history that half the world's population lived in countries holding national elections within a calendar year.[9] In January 2024, the World Economic Forum reported that over a thousand experts had identified misinformation and disinformation as the leading risk to global order over the next two years — ahead of climate change and war.[10]

---

[6] This article analyses laws based on a survey of 193 member states of the United Nations, with the number of documents increasing from 45 in 2016 to 151 in 2023. In 2016, there were 37 states with such laws, growing to 102 in 2023. The raw data is available as an appendix. *See* the discussion of data collection in part three.

[7] Editorial, *Cambridge Analytica Controversy Must Spur Researchers to Update Data Ethics*, 555 NATURE 559 (2018); Jacob L. Nelson and Harsh Taneja, *The Small, Disloyal Fake News Audience: The Role of Audience Availability in Fake News Consumption*, 20(10) NEW MEDIA & SOCIETY 3720 (2018); JAMES W. CORTADA AND WILLIAM ASPRAY, FAKE NEWS NATION: THE LONG HISTORY OF LIES AND MISINTERPRETATIONS IN AMERICA (2019); Margaret Hu, *Cambridge Analytica's Black Box*, 7(2) BIG DATA & SOC'Y (2020); JIM MACNAMARA, BEYOND POST-COMMUNICATION: CHALLENGING DISINFORMATION, DECEPTION, AND MANIPULATION (2020); IMKE HENKEL, DESTRUCTIVE STORYTELLING: DISINFORMATION AND THE EUROSCEPTIC MYTH THAT SHAPED BREXIT (2021); NEWTON LEE, FACEBOOK NATION: TOTAL INFORMATION AWARENESS (3rd ed. 2021).

[8] *See, e.g.,* João Marecos et al., *Health Misinformation and Freedom of Expression: Considerations for Policymakers*, 18 HEALTH ECONOMICS, POLICY AND LAW 204 (2023).

[9] *See, e.g.,* Stefan Feuerriegel et al., *Research Can Help to Tackle AI-generated Disinformation*, 7(11) NATURE HUMAN BEHAVIOUR 1818 (2023); Danielle Allen and E. Glen Weyl, *The Real Dangers of Generative AI*, 35(1) JOURNAL OF DEMOCRACY 147 (2024). On the relationship between truth and democracy, *see* JÜRGEN HABERMAS, BETWEEN NATURALISM AND RELIGION: PHILOSOPHICAL ESSAYS 143-44 (2008).

[10] The Global Risks Report 2024 (World Economic Forum, Geneva, January 2024), *at* https://www3.weforum.org/docs/WEF_The_Global_Risks_Report_2024.pdf, at 8. On a ten year horizon, environmental concerns ranked first, but misinformation and disinformation still came ahead of the threat posed by interstate armed conflict or 'adverse outcomes of AI technologies'.

This article will first survey efforts to understand the problem and its origin. The mere existence of information that is not true is not a harm in itself. In general, it is the sharing of such information and the impact on society that may constitute a harm worthy of regulatory action, whether to protect vulnerable individuals, certain public institutions, and (perhaps) the bonds of trust that underpin much of society. Part two considers different possible responses and the risks associated with them. Such efforts face challenges, particularly if they limit access to information through censorship. In the context of larger debates over the governance of AI, regulators across the globe are struggling to address perceived harms associated with generative AI while not unduly limiting innovation or driving it elsewhere. Regulation is understood here to include rules, standards, and less formal forms of supervised self-regulation. Policy interventions are broader still, including educational and social policies intended to build resilience among citizens and users.[11]

The present article focuses on legislation, and part three presents the findings of a survey of legislative activity worldwide.[12] Among the findings are that laws tended to be introduced first in countries that were less free, in terms of civil liberties, and poorer, as measured by GDP per capita. Internet penetration does not seem to have been a driving factor. African and Asian countries were the first to adopt such laws in significant numbers. More recently, Asian states are responsible for a significant increase in laws and have tended to grant greater powers to governments. The focus of such laws is most frequently on national security broadly construed, though 2020 saw a spike in laws addressing public health. Unsurprisingly, governments with fewer legal constraints on government action have generally adopted more robust positions in dealing with false information. Despite early reservations, however, growth in such laws is now steepest in Western states. Though there are diverse views on the appropriate response to false information online, the need for legislation of some kind appears now to be global.

---

[11] James Meese and Edward Hurcombe, Regulating Misinformation (RMIT, Melbourne, 2020), *at* https://apo.org.au/node/309357.

[12] This article is explicitly an examination of 'law in the books' rather than 'law in action', focusing on the provisions adopted by legislative bodies rather than their implementation and effectiveness. Such an empirical overview of form may lay the groundwork for subsequent work on the impact of such laws over time. Cf Roscoe Pound, *Law in Books and Law in Action*, 44 Am. L. Rev. 12 (1910).

# 1   Lawful but Awful Content

The focus of this article is efforts to regulate online speech that does not fall into traditional categories subject to prohibition or regulation. Sometimes termed 'lawful but awful', this comprises potentially harmful content that is not covered by laws on obscenity, incitement, harassment, defamation, national security laws, and the like.

It is helpful to distinguish between three types of content that might be the target of regulation, based on veracity and the intention of the creator or sharer. 'Misinformation' is information that is false or misleading but does not have malicious intent. It may include urban legends or rumours. 'Disinformation' is the deceptive use of false information. It may be used to push an agenda or manipulate a group. A third category of 'mal-information' aims to inflict personal harm and includes prejudice, hate speech, up to harassment, even if it relies on information that is in some sense true.[13] These can be represented in a two-by-two matrix as shown in Table 1.

|  |  | *Veracity* | |
|---|---|---|---|
|  |  | *True* | *False* |
| *Intent* | *Benevolent/neutral* | - | Misinformation |
|  | *Malevolent* | Mal-information | Disinformation |

*Table 1: Categorization of potentially harmful information*

These distinctions are important because the information itself is not typically the problem, but how it affects the real world. The mere fact of a statement being untrue should not make it a target of concern and highlights the risk of over-inclusion in any response. This was graphically on display when Singapore's Media Literacy Council engaged in an ill-conceived public education campaign that erroneously included 'satire' as an example of fake news that was to be targeted by new legislation. (It issued a correction.[14]) Harms that may be worthy of concern include the impact on vulnerable individuals and certain public institutions; arguably, they may also encompass the bonds of trust that hold society together. This part offers a brief survey of such considerations, before turning in part two to the possible responses and then in part three to the emerging patterns in legislative behaviour.

---

[13] *See* generally Claire Wardle and Hossein Derakhshan, Information Disorder: Toward an Interdisciplinary Framework for Research and Policy Making (Council of Europe, DGI(2017)09, September 2017), *at* https://rm.coe.int/090000168076277c.

[14] Tee Zhuo, *Media Literacy Council Apologises for Facebook Post on Satire Being Fake News*, STRAITS TIMES, 8 September 2019. *See also* Jwen Fai Low et al., *Distinguishing Between Fake News and Satire with Transformers*, 187 EXPERT SYSTEMS WITH APPLICATIONS 115824 (2022); Fernando Miró-Llinares and Jesús C. Aguerri, *Misinformation About Fake News: A Systematic Critical Review of Empirical Studies on the Phenomenon and Its Status as a "Threat"*, 20(1) EUROPEAN JOURNAL OF CRIMINOLOGY 356 (2023).

## 1.1 Vulnerable Individuals

The use of online tools to target individuals for the purposes of fraud, for example, appears to be on the rise.[15] This follows an increase in online fraud generally, partly due to the overall increase in online transactions — especially during the pandemic, when entrepreneurial criminals developed novel means of defrauding victims while also observing social distancing.[16] Such activities typically have criminal penalties or civil remedies available, though there are practical challenges to identifying perpetrators.[17] The gap to be addressed includes adjacent behaviour that is socially harmful but not rising to the level of an existing crime or civil wrong.

More recently, generative AI has significantly lowered the costs of manufacturing information that is untrue or distorts the truth, with significant attention being given to the rise of 'deepfakes': images or videos in which a person's face or body is altered, often with malicious intent. Though the focus is normally on deepfakes of famous people — distorted for pornographic,[18] political,[19] or financial purposes,[20] — near-term threats include generative AI being used for phishing attacks (tricking users into revealing confidential information) or extortion targeting the wider public.[21] This might take the form of a phone call or even a video from a loved one that is faked. Savvy users might no longer be willing to help that Nigerian prince get access to his millions,[22] but who among us would not assist a child or spouse in a panic?[23]

---

[15] INTERPOL Financial Fraud Assessment: A Global Threat Boosted by Technology (Interpol, Lyon, 11 March 2024), *at* https://www.interpol.int/en/News-and-Events/News/2024/INTERPOL-Financial-Fraud-assessment-A-global-threat-boosted-by-technology.

[16] Jay P. Kennedy, Melissa Rorie, and Michael L. Benson, *COVID-19 Frauds: An Exploratory Study of Victimization During a Global Crisis*, 20(3) CRIMINOLOGY & PUBLIC POLICY 493 (2021).

[17] Cassandra Cross, *"Oh We Can't Actually do Anything About That": The Problematic Nature of Jurisdiction for Online Fraud Victims*, 20(3) CRIMINOLOGY & CRIMINAL JUSTICE 358 (2020).

[18] Anne Pechenik Gieseke, *"The New Weapon of Choice": Law's Current Inability to Properly Address Deepfake Pornography*, 73(5) VAND. L. REV. 1479 (2020); Dean Fido, Jaya Rao, and Craig A. Harper, *Celebrity Status, Sex, and Variation in Psychopathy Predicts Judgements of and Proclivity to Generate and Distribute Deepfake Pornography*, 129 COMPUTERS IN HUMAN BEHAVIOR 107141 (2022).

[19] John Twomey et al., *Do Deepfake Videos Undermine Our Epistemic Trust? A Thematic Analysis of Tweets that Discuss Deepfakes in the Russian Invasion of Ukraine*, 18(10) PLOS ONE e0291668 (2023).

[20] Nir Kshetri, *The Economics of Deepfakes*, 56(8) COMPUTER 89 (2023).

Another prominent example of harm to individuals is so-called 'revenge porn', in which potentially genuine information (intimate photographs, for example) is used without consent in order to tarnish a reputation or otherwise harm an individual. In addition to general obscenity laws, specific legislation has been adopted in several jurisdictions to address such activities.[24]

Protection of children online has exercised regulators in many jurisdictions. Despite notional age requirements imposed by laws such as the US Children's Online Privacy Protection Act 1998,[25] such mechanisms are trivial to evade.[26] Particular attention has been given to laws protecting children from sexual exploitation and abuse, which mostly fall outside the scope of this study.[27] Nonetheless, some laws have sought to supplement such provisions to reduce the risk of online grooming.[28]

---

[21] Dan Milmo, *Company Worker in Hong Kong Pays Out £20m in Deepfake Video Call Scam*, GUARDIAN, 5 February 2024. *See* generally Fakhar Abbas and Araz Taeihagh, *Unmasking Deepfakes: A Systematic Review of Deepfake Detection and Generation Techniques Using Artificial Intelligence*, 252 EXPERT SYSTEMS WITH APPLICATIONS 124260 (2024).

[22] Ojeifoh Okosun and Uchenna Ilo, *The Evolution of the Nigerian Prince Scam*, 30(6) JOURNAL OF FINANCIAL CRIME 1653 (2023).

[23] Charles Bethea, *The Terrifying A.I. Scam That Uses Your Loved One's Voice*, NEW YORKER, 7 March 2024.

[24] *See, e.g.,* Emma Burnett, *Criminalizing Nonconsensual Pornography*, 103(6) BOSTON UNIVERSITY LAW REVIEW 1819 (2023); Karolina Mania, *Legal Protection of Revenge and Deepfake Porn Victims in the European Union: Findings From a Comparative Legal Study*, 25(1) TRAUMA, VIOLENCE, & ABUSE 117 (2024).

[25] Children's Online Privacy Protection Act (COPPA) 1998 (U.S.).

[26] Brian O'Neill, *Who Cares? Practical Ethics and the Problem of Underage Users on Social Networking Sites*, 15(4) ETHICS AND INFORMATION TECHNOLOGY 253 (2013); Mariya Stoilova, Monica Bulger, and Sonia Livingstone, *Do Parental Control Tools Fulfil Family Expectations for Child Protection? A Rapid Evidence Review of the Contexts and Outcomes of Use*, 18(1) JOURNAL OF CHILDREN AND MEDIA 29 (2024).

[27] *See, e.g.,* Amy, Vicky, and Andy Child Pornography Victim Assistance Act 2018 (U.S.).

[28] *See, e.g.,* Sexual Offences Act 2003 (UK), s15A. Jordan Hill, Policy Responses to False and Misleading Digital Content: A Snapshot of Children's Media Literacy (OECD, Paris, OECD Education Working Papers No. 275, 2022), *at* https://www.oecd-ilibrary.org/docserver/1104143e-en.pdf.

More generally, online content that encourages self-harm, promotes or glorifies violence, or encourages and entrenches societal divisions may not rise to the level of incitement or hate speech but be the subject of regulation seeking to encourage or compel content moderation on the part of platforms.[29]

In addition to increasing the opportunity for online fraud, the Covid-19 pandemic also highlighted the cost of distorted public health messages. Though sometimes a result of active disinformation campaigns,[30] much of this was more properly understood as misinformation — ill-informed but not malicious sharing of gossip, rumours, and the speculations of those who had 'done their own research'.[31] Extreme cases of misinformation included dangerous health advice, such as ingesting bleach or using unapproved medication for treatment,[32] though more widespread dismissal of the seriousness of Covid-19 or exaggeration of the risks of vaccines increased the death toll in some locations.[33] Health misinformation also appears to have fuelled anti-vaccine movements, with wider declines in vaccination acceptance rates. Deaths due to measles have increased 43 percent worldwide, for example, more than doubling in the United States.[34]

---

[29] *See, e.g.,* Online Safety Act 2023 (UK).

[30] *See, e.g.,* Sara Monaci and Simone Persico, *The COVID-19 Vaccination Campaign and Disinformation on Twitter: The Role of Opinion Leaders and Political Social Media Influencers in the Italian Debate on Green Pass*, 16 INT'L J. COMM. 5885 (2022).

[31] MANUFACTURING GOVERNMENT COMMUNICATION ON COVID-19: A COMPARATIVE PERSPECTIVE (Philippe J. Maarek ed., 2022); J. Agley and Y. Xiao, *Low Trust in Science May Foster Belief in Misinformation by Aligning Scientifically Supported and Unsupported Statements*, 143(4) PERSPECTIVES IN PUBLIC HEALTH 199 (2023); Neil Levy, *Echoes of Covid Misinformation*, 36(5) PHILOSOPHICAL PSYCHOLOGY 931 (2023).

[32] Meghan A. Cook and Nicholas Brooke, *Event-Based Surveillance of Poisonings and Potentially Hazardous Exposures over 12 Months of the COVID-19 Pandemic*, 18(21) INTERNATIONAL JOURNAL OF ENVIRONMENTAL RESEARCH AND PUBLIC HEALTH 11133 (2021).

[33] Kris Hartley and Minh Khuong Vu, *Fighting Fake News in the COVID-19 Era: Policy Insights from an Equilibrium Model*, 53(4) POLICY SCIENCES 735 (2020); PSYCHOLOGICAL INSIGHTS FOR UNDERSTANDING COVID-19 AND MEDIA AND TECHNOLOGY (Ciarán Mc Mahon ed., 2020); BARRIE GUNTER, PSYCHOLOGICAL INSIGHTS ON THE ROLE AND IMPACT OF THE MEDIA DURING THE PANDEMIC: LESSONS FROM COVID-19 (2022); Sun Kyong Lee et al., *Misinformation of COVID-19 Vaccines and Vaccine Hesitancy*, 12(1) SCIENTIFIC REPORTS 13681 (2022/08/11 2022); Peter Hotez, *Anti-Science Conspiracies Pose New Threats to US Biomedicine in 2023*, 5(6) FASEB BIOADVANCES 228 (Jun 2023

2023-06-06 2023).

[34] Helen Bedford and David Elliman, *Measles Rates Are Rising Again*, 384 BMJ q259 (2024); Riis Williams, *What to Know about Measles Outbreaks in the US*, SCIENTIFIC AMERICAN, 22 March 2024.

## 1.2 Public Institutions

Public health is an example of a situation in which undermining official messages may be seen as a harm worthy of government intervention. To what extent should such interventions protect the legitimacy and reputation of public institutions more generally?

Following the Cambridge Analytica scandal of 2016[35] and challenges to the integrity of the 2020 US presidential election,[36] a significant focus has been on safeguarding electoral processes from undue interference.[37] The rise of AI-generated content in campaigns is well documented,[38] though the actual impact on results is unclear. Argentina's 2023 ballot was widely described as the first 'AI election',[39] though it does not appear to have affected the outcome significantly. Robocalls using a cloned version of US President Joe Biden's voice during the New Hampshire primary in January 2024 advised voters to stay home, leading to a hasty effort by the Federal Communications Commission to ban such disinformation.[40] Several other countries have adopted laws specifically focused on elections.[41]

---

[35] *See supra* n 7.

[36] Samia Benaissa Pedriza, *Sources, Channels and Strategies of Disinformation in the 2020 US Election: Social Networks, Traditional Media, and Political Candidates*, 2(4) JOURNALISM AND MEDIA 605 (2021).

[37] *See* GUILLERMO LOPEZ-GARCIA et al., POLITICS OF DISINFORMATION: THE INFLUENCE OF FAKE NEWS ON THE PUBLIC SPHERE (2021); THE EPISTEMOLOGY OF DECEIT IN A POSTDIGITAL ERA: DUPERY BY DESIGN (Alison MacKenzie, Jennifer Rose, and Ibrar Bhatt eds., 2021); ASSANE DIAGNE et al., MISINFORMATION POLICY IN SUB-SAHARAN AFRICA: FROM LAWS AND REGULATIONS TO MEDIA LITERACY (2021); HERMAN WASSERMAN AND DANI MADRID-MORALES, DISINFORMATION IN THE GLOBAL SOUTH (2022); JAMES GOMEZ AND ROBIN RAMCHARAN, FAKE NEWS AND ELECTIONS IN SOUTHEAST ASIA: IMPACT ON DEMOCRACY AND HUMAN RIGHTS (2023).

[38] Rumman Chowdhury, *AI-Fuelled Election Campaigns Are Here — Where Are the Rules?*, 628 NATURE 237 (2024).

[39] Jack Nicas and Lucía Cholakian Herrera, *Is Argentina the First A.I. Election?*, N.Y. TIMES, 15 November 2023.

[40] Julia Shapero, *FCC Targets AI-generated Robocalls After Biden Primary Deepfake*, THE HILL, 1 February 2024.

[41] *See, e.g.,* Public Offices Election Act 2019 (Japan), s 235. *See also* Tom Dobber et al., *Do (Microtargeted) Deepfakes Have Real Effects on Political Attitudes?*, 26(1) THE INTERNATIONAL JOURNAL OF PRESS/POLITICS 69 (2021); Michael Hameleers, Toni G. L. A. van der Meer, and Tom Dobber, *You Won't Believe What They Just Said! The Effects of Political Deepfakes Embedded as Vox Populi on Social Media*, 8(3) SOCIAL MEDIA + SOCIETY (2022); Gene M. Grossman and Elhanan Helpman, *Electoral Competition with Fake News*, 77 EUROPEAN JOURNAL OF POLITICAL ECONOMY 102315 (2023); Prerna Juneja, Md Momen Bhuiyan, and Tanushree Mitra, *Assessing Enactment of Content Regulation Policies: A post hoc Crowd-Sourced Audit of Election Misinformation on YouTube*, 545 PROCEEDINGS OF THE 2023 CHI CONFERENCE ON HUMAN FACTORS IN COMPUTING SYSTEMS (CHI '23) (2023).

Larger efforts to protect governments themselves run into the question of whether the intent is to safeguard public institutions or the political fortunes of the government of the day. This is a familiar problem in national security law, where the mandates of agencies with domestic and international remits are often framed in general terms and open to wide interpretation. In Britain, for example, the Security Service (better known as MI5) operated for most of the latter part of the twentieth century on the basis of a six-paragraph administrative direction that described its role as 'the Defence of the Realm as a whole'.[42] The Secret Intelligence Service (scilicet MI6) was only officially acknowledged even to exist in 1992 — well after the release of the sixteenth James Bond film portraying the exploits of its most famous fictional agent.[43] Efforts to rein in intelligence agencies — in the US context referred to around the Watergate period as 'rogue elephants'[44] — played out in debates over how broadly or narrowly to construe their mandates, with the subtext often being that in the absence of a strong rule of law foundation, they were open to abuse either at the hands of unscrupulous political leaders or entrepreneurial agency directors.[45]

Nonetheless, national security concerns are among the most frequently cited justifications for laws regulating online speech.[46] 'Hostile information campaigns' have become both a new tool in the arsenal of intelligence services and a reason for cracking down on destabilising (or inconvenient) content.[47] In some ways, this is merely an extension of the role propaganda has always played in international affairs, though new online tools have

---

[42] Maxwell-Fyfe Directive (issued by the UK Home Secretary, Sir David Maxwell-Fyfe, to the Director-General MI5, 1952), reprinted in LAURENCE LUSTGARTEN AND IAN LEIGH, IN FROM THE COLD: NATIONAL SECURITY AND PARLIAMENTARY DEMOCRACY 517 (1994).

[43] SIMON CHESTERMAN, ONE NATION UNDER SURVEILLANCE: A NEW SOCIAL CONTRACT TO DEFEND FREEDOM WITHOUT SACRIFICING LIBERTY 132 (2011).

[44] Marina Caparini, *Controlling and Overseeing Intelligence Services in Democratic States*, in DEMOCRATIC CONTROL OF INTELLIGENCE SERVICES: CONTAINING ROGUE ELEPHANTS 3 (Hans Born and Marina Caparini eds., 2007).

[45] *See* CHESTERMAN, *supra* note 43, at 138-41. *See also* The Johannesburg Principles on National Security, Freedom of Expression and Access to Information (Article 19, Johannesburg, November 1996), *at* https://www.article19.org/wp-content/uploads/2018/02/joburg-principles.pdf.

[46] *See* part three.

[47] Thomas Paterson and Lauren Hanley, *Political Warfare in the Digital Age: Cyber Subversion, Information Operations and "Deep Fakes"*, 74(4) AUSTRALIAN JOURNAL OF INTERNATIONAL AFFAIRS 439 (2020); Gururaghav Raman et al., *How Weaponizing Disinformation Can Bring Down a City's Power Grid*, 15(8) PLOS ONE e0236517 (2020); Spencer McKay and Chris Tenove, *Disinformation as a Threat to Deliberative Democracy*, 74(3) POLITICAL RESEARCH QUARTERLY 703 (2021); Deen Freelon et al., *Black Trolls Matter: Racial and Ideological Asymmetries in Social Media Disinformation*, 40(3) SOCIAL SCIENCE COMPUTER REVIEW 560 (2022).

enabled campaigns to be waged at speed and at scale, as well as by means that can be disavowed.[48]

In addition to electoral processes, judicial and financial institutions are among other public functions sometimes seen as deserving of special protection. The judiciary is unable to defend itself as robustly as other more political arms of government, with old measures such as scandalizing the judiciary being revived in some jurisdictions in the name of protecting the rule of law. Financial bodies have long been subject to efforts at manipulation, with online tools challenging the ability of traditional measures to respond in time to prevent harms.[49] In 2010, the 'Flash Crash' — in which high-frequency trading algorithms wiped a trillion dollars off the New York Stock Exchange in half an hour — led to the introduction of circuit breakers to slow trading in the face of anomalous activity.[50] Similar measures may be required to deal with misinformation or disinformation intended to manipulate prices.[51]

## 1.3 The Liar's Dividend

A more amorphous form of harm may be the erosion of faith in the notion of truth itself. Though not typically the subject of legislation, concern has been expressed by commentators,[52] governments,[53] and international organizations[54] at the more general impact of proliferating 'fake news'. In such an environment, in addition to specific deepfakes gaining traction, a perverse product is the 'liar's dividend' — where the public dismiss genuine scandals or allegations of wrongdoing because the basis of truth or falsity has become so muddied and confused.[55] It is common in this context to focus on the *quality* of disinformation, though its *quantity* may be the bigger concern. If synthetic content ends up flooding the Internet, the result may not be that people believe the lies, but that they cease to believe anything at all. Any uncomfortable or inconvenient information will be dismissed as 'fake', while many will credulously accept data that reinforce their own worldview as 'true'.[56]

---

[48] Lynnette H.X. Ng and Araz Taeihagh, *How Does Fake News Spread? Understanding Pathways of Disinformation Spread Through APIs*, 13(4) POL'Y & INTERNET 560 (2021).

[49] Viktor Manahov, *Cryptocurrency Liquidity During Extreme Price Movements: Is There a Problem with Virtual Money?*, 21(2) QUANTITATIVE FINANCE 341 (2021).

[50] *See* E. Wes Bethel et al., *Federal Market Information Technology in the Post Flash Crash Era: Roles for Supercomputing*, 7(2) THE JOURNAL OF TRADING 9 (2012); MICHAEL LEWIS, FLASH BOYS: A WALL STREET REVOLT (2014); Andrei Kirilenko et al., *The Flash Crash: High-Frequency Trading in an Electronic Market*, 72 J. FIN. 967 (2017); Simon Chesterman, *"Move Fast and Break Things": Law, Technology, and the Problem of Speed*, 33 SINGAPORE ACADEMY OF LAW JOURNAL 5 (2021).

[51] Cf Tom C.W. Lin, *The New Market Manipulation*, 66(6) EMORY L.J. 1253, 1292-94 (2017).

[52] *See, e.g.,* JOHN JONES AND MICHAEL TRICE, PLATFORMS, PROTESTS, AND THE CHALLENGE OF NETWORKED DEMOCRACY (2020); THE EPISTEMOLOGY OF FAKE NEWS (Sven Bernecker, Amy K. Flowerree, and Thomas Grundmann eds., 2021); POLITICAL EPISTEMOLOGY (Elizabeth Edenberg and Michael Hannon eds., 2021); Daniel Cassar, *The Misinformation Threat: A Techno-Governance Approach for Curbing the Fake News of Tomorrow*, 4(4) DIGITAL GOVERNMENT: RESEARCH AND PRACTICE 1 (2023).

[53] Yuen-C Tham, *Fake News Law Necessary to Prevent Crisis of Trust that Has Hit Other Countries, Says Shanmugam*, STRAITS TIMES, 9 May 2019.

A related concern is the way in which most of us access information online. For the past two decades, 'Googling' meant entering specific search terms to obtain a list of results.[57] The search terms reminded us that we were interacting with a program and the list of results highlighted the multiple possible answers, of which a discreet amount of advertising disclosed how the service was paid for. The move to generative AI chatbots like ChatGPT, Gemini, Claude, and the like have now raised the prospect of such searches being replaced by natural language queries of that will yield a single response. It seems highly likely that many will accept the responses of such intelligent agents as 'truthful enough', with the additional possibility that, as these chatbots learn our preferences (and prejudices), they may serve to reinforce them in an algorithmic echo chamber.[58] This is a known problem in generative AI, where fine-tuning may lead to a trade-off between helpfulness and truthfulness — a phenomenon known as sycophancy.[59]

---

[54] *See, e.g.,* Journalism, 'Fake News' and Disinformation: A Handbook for Journalism Education and Training (UNESCO, Paris, 2018), *at* https://en.unesco.org/fightfakenews; Misinformation and Disinformation: An International Effort Using Behavioural Science to Tackle the Spread of Misinformation (OECD, Paris, OECD Public Governance Policy Papers, No. 21, 2022), *at* https://www.oecd-ilibrary.org/content/paper/b7709d4f-en; Facts not Fakes: Tackling Disinformation, Strengthening Information Integrity (OECD, Paris, 2024), *at* https://www.oecd.org/governance/facts-not-fakes-tackling-disinformation-strengthening-information-integrity-d909ff7a-en.htm.

[55] Bobby Chesney and Danielle Citron, *Deep Fakes: A Looming Challenge for Privacy, Democracy, and National Security*, 107(6) CAL. L. REV. 1753, 1785-86 (2019).

[56] *See also* M. Anne Britt et al., *A Reasoned Approach to Dealing With Fake News*, 6(1) POLICY INSIGHTS FROM THE BEHAVIORAL AND BRAIN SCIENCES 94 (2019); Chau Tong et al., *"Fake News Is Anything They Say!" — Conceptualization and Weaponization of Fake News among the American Public*, 23(5) MASS COMMUNICATION & SOCIETY 755 (2020); Valerie F. Reyna, *A Scientific Theory of Gist Communication and Misinformation Resistance, with Implications for Health, Education, and Policy*, 118(15) PROCEEDINGS OF THE NATIONAL ACADEMY OF SCIENCES - PNAS 1 (2021); ROB COVER, ASHLEIGH HAW, AND JAY DANIEL THOMPSON, FAKE NEWS IN DIGITAL CULTURES: TECHNOLOGY, POPULISM AND DIGITAL MISINFORMATION (2022); Yinuo Geng, *Comparing "Deepfake" Regulatory Regimes in the United States, the European Union, and China*, 7 GEORGETOWN LAW TECHNOLOGY REVIEW 157 (2023); CAUSES AND SYMPTOMS OF SOCIO-CULTURAL POLARIZATION: ROLE OF INFORMATION AND COMMUNICATION TECHNOLOGIES (Israr Qureshi et al. eds., 2022).

[57] ROSIE GRAHAM, INVESTIGATING GOOGLE'S SEARCH ENGINE: ETHICS, ALGORITHMS, AND THE MACHINES BUILT TO READ US (2023).

## 2 A Toolbox of Responses

What, if anything, should governments do in response to these perceived threats? For much of the history of the internet, the answer was very little. Indeed, the United States broadly embraced this position in section 230 of the 1996 Communications Decency Act, which absolved Internet platforms of responsibility for the content posted on them.[60] Other jurisdictions struggled to apply existing rules to the online medium, notably pornography and hate speech,[61] much as they also struggled to enforce copyright on digital platforms.[62]

In practice, much depended on technical interventions by the platforms themselves, with limited success. Efforts to regulate any aspect of the digital information pipeline have typically faced considerable resistance, either on the basis that it would reduce the benefits of connectivity or inappropriately censor content. It also runs contrary to vested interests of corporate actors, which have lobbied against regulation. In the context of larger debates over the governance of AI, regulators across the globe are today struggling to address perceived harms associated with generative AI while not unduly limiting innovation or driving it elsewhere.[63]

---

[58] *See, e.g.,* Scott Monteith et al., *Artificial Intelligence and Increasing Misinformation*, 224(2) BRITISH JOURNAL OF PSYCHIATRY 33 (2024). Cf Corinne Tan, *The Curious Case of Regulating False News on Google*, 46 COMPUTER L. & SECURITY REP. 105738 (2022) (discussing earlier trends in this direction).

[59] Mrinank Sharma et al., Towards Understanding Sycophancy in Language Models (Cornell University Library, arXiv.org, Ithaca, 2023), *at* https://go.exlibris.link/ZXDvbTn1.

[60] Communications Decency Act 1996 (US), 47 US Code §230. *See also* Cheng-Chi Chang, *Revisiting Disinformation Laws in the Age of Social Media*, 6(4) ARIZONA LAW JOURNAL OF EMERGING TECHNOLOGIES (2023).

[61] *See, e.g.,* Corinne Tan, *Regulating Disinformation on Twitter and Facebook*, 31(4) GRIFFITH LAW REVIEW 513 (2022).

[62] David Tan, *Fair Use and Transformative Play in the Digital Age, in* RESEARCH HANDBOOK ON INTELLECTUAL PROPERTY IN MEDIA AND ENTERTAINMENT 102 (Megan Richardson and Sam Ricketson eds., 2017); Simon Chesterman, *Good Models Borrow, Great Models Steal: Intellectual Property Rights and Generative AI*, puae006 POLICY AND SOCIETY forthcoming (2024).

The starting point is to be clear about what the objectives are and the tools and levers available. Spreading malicious content is already the subject of regulation in many jurisdictions. Though there is wariness about unnecessary limits on freedom of speech, even in broadly libertarian jurisdictions like the United States one is not allowed to yell 'fire!' in a crowded theatre. Here the responses will be divided into potential regulation of the production, distribution, and consumption of problematic content.

## 2.1 Production

Generative AI has raised the question of whether the tools that generate content should themselves be regulated. Private activity is not normally regulated in this way — a hateful lie written in my personal diary is not a crime; nor do we generally prohibit unpublished threats typed on a word processor.

A notable exception is that several jurisdictions now make it an offence to create or possess child pornography, including synthetic images in which no actual child was harmed and even if the images are not shared.[64] Photographs of a wrong, such as abuse of children or acts of violence, are generally punishable because of the underlying harm. A *simulated* wrong — in which the underlying harm did not take place — may not be prohibited. The US Supreme Court, for example, has struck down provisions of the Child Pornography Prevention Act of 1996 that would have criminalized such 'speech' on the basis that it 'records no crime and creates no victims by its production'.[65]

Different approaches to synthetic pornography involving adults only have been followed in various jurisdictions, but the mere creation of such material is rarely prohibited, in part because of the difficulty of enforcing any such laws.[66] In 2024, Britain's Ministry of Justice announced that the creation of a sexually explicit 'deepfake' image would be made an offence, punishable by a fine even if there was no intent to share it. (In the event that such an image is shared, penalties include jailtime under existing provisions in the Online Safety Act.)[67]

---

[63] *See, e.g.,* Chesterman, *supra* note 62.

[64] *See, e.g.,* Dominique Moritz, Ashley Pearson, and Larissa S. Christensen, *Exploring Virtual Child Sexual Abuse Material Law: When Creativity Is Criminalised*, 24(4) MEDIA AND ARTS LAW REVIEW 255 (2022). This is in addition to specific cases in which repeat offenders have been banned from using tools to create such images: Shanti Das, *Sex Offender Banned From Using AI Tools in Landmark UK Case*, GUARDIAN, 21 April 2024.

[65] *Ashcroft v. Free Speech Coalition*, 535 U.S. 234, 250 (2002).

[66] Cf Dan Milmo, *OpenAI Considers Allowing Users to Create AI-generated Pornography*, GUARDIAN, 9 May 2024.

## 2.2 Distribution

For the most part, the harm is in the impact the information has on other users and society. In addition to punishing those who intend harms such as fraud, hate speech, or defamation, attention has episodically focused on the responsibility of platforms that host and facilitate access. Alternative approaches are also possible, sometimes pushed by the platforms themselves. These include focusing on the speed with which potentially harmful content can spread, or ensuring the traceability of content and limiting the ability of those who produce harmful content to hide behind the anonymity of the internet. More recently, attention has turned to the need to identify synthetic content created by generative AI and distinguish it from 'real' or human-generated content.

**Platform Responsibility**

As indicated earlier, the United States took the early position that platforms should not generally be held responsible for the content posted.[68] This legislative intervention followed lawsuits against online discussion platforms that turned on whether internet service providers were distributors of content analogous to telephone companies, who have limited control over the words said on their lines, or publishers akin to newspapers, who bear greater responsibility for content. Section 230 of the 1996 Communications Decency Act provided that 'No provider or user of an interactive computer service shall be treated as the publisher or speaker of any information provided by another information content provider.'[69] The same provision also protected 'Good Samaritan' blocking and screening of material that the 'provider or user considers to be obscene, lewd, lascivious, filthy, excessively violent, harassing, or otherwise objectionable'.[70] Though there were subsequent efforts at additional moderation by platforms at the state level, prompting several constitutional challenges, it is striking that no major technology regulation was subsequently passed at the Federal level by the United States until its attempt to force the divestiture of TikTok from its Chinese parent company ByteDance in 2024.[71]

---

[67] Criminal Justice Bill (2024), details at https://bills.parliament.uk/bills/3511. *See Creating Sexually Explicit Deepfake Images to Be Made Offence in UK*, GUARDIAN, 16 April 2024.

[68] *See supra* n 60.

[69] 47 US Code §230(c)(1).

[70] 47 US Code §230(c)(2)(A).

In the same year that the United States guaranteed platform neutrality (1996), laws concerning online material were also adopted in Eritrea and the Russian Federation with somewhat different objectives. Eritrea's law updated its press regulation to cover material 'disseminated to the public by means of new techniques, technology and others'.[72] It provided for freedom of the press and rights of journalists, but also stated that journalists 'may neither disseminate information the veracity of which has not been ascertained nor distort information.'[73] Other provisions require the correction of 'inaccurate news or information' and potential fines for 'wrong or false news'.[74] Russia's amendments to its penal code were part of a larger effort at modernization, including a new category of crimes related to computer information. The crime of disseminating knowingly false fabrications that 'defame the Soviet state' was removed, but narrower prohibitions were added concerning knowingly false communications about acts of terrorism, crimes, and elections.[75] Other early actors in the late 1990s were Azerbaijan and Egypt. The former adopted a law on information, informatization, and information protection;[76] the latter amended its criminal

---

[71] 21st Century Peace Through Strength Act 2024 (US).

[72] Press Proclamation No. 90/1996 1996 (Eritrea), s 3.

[73] *Id.*, s 5(2)(d).

[74] *Id.*, ss 11, 15. *See also* Kjetil Tronvoll and Daniel R. Mekonnen, The African Garrison State: Human Rights and Political Development in Eritrea (2014).

[75] The Criminal Code of the Russian Federation, No. 63-FZ 1996 (Russian Federation), arts 207, 306, 142; Anatolyi V. Naumov, *The New Russian Criminal Code as a Reflection of Ongoing Reforms*, 8(2) Crim. L.F. 191 (1997

2023-11-22 1997); Christian Caryl, *Big Brother Covets All the E-mail*, US News & World Report, 14 September 1998; William Burnham, *The New Russian Criminal Code: A Window onto Democratic Russia*, 26(4) Review of Central and East European Law 365 (2000). *See* now Liudmila Sivetc, *State Regulation of Online Speech in Russia: The Role of Internet Infrastructure Owners*, 27 Int'l J.L. & Info. Tech. 28 (2019).

[76] İnformasiya, informasiyalaşdırma və informasiyanın mühafizəsi haqqında [About information, informatization and information protection} 1998 (Azerbaijan); Jean Patterson, *Azerbaijan on the Internet*, 6(4) Azerbaijan International 60 (1998).

code to prohibit disclosing false or tendentious news, damaging public or national interests.[77]

These distinct approaches — wariness of the potential impact of false information spreading quickly through a new medium, versus a desire to protect and promote the very actors who were making that new medium possible — set the tone for the early years of internet regulation. The general wariness of overregulating platforms continued, particularly in developed countries with constitutional and other protections of free speech.[78] Those laws that were adopted tended to target specific challenges such as spam,[79] children,[80] and cyberbullying.[81] Germany was an outlier among developed countries with its 2017 Network Enforcement Act (NetzDG),[82] later credited with marking a shift in the regulatory approach of European and other Western countries.[83]

More robust laws are, by contrast, common in countries with fewer protections of civil liberties. In 2019, for example, Singapore adopted the Protection from Online Falsehoods and Manipulation Act (POFMA), which empowers ministers to make correction directions for false statements of fact if, in the minister's opinion, it is in the public interest to do so.[84] This typically takes the form of an additional link to a government website, though not requiring removal of the original falsehood.[85] At the time, Singapore was criticised for the restrictions this imposed on speech and the manner in which it might be used to silence dissent.[86] Despite any such reservations, in the wake of the Covid-19 pandemic of 2020, there was a surge in efforts to legislate.[87]

---

[77] Law No. 58 of 1937 Promulgating the Penal Code 1937 (Egypt), art 80C, 102*bis*.

[78] Cf Barrie Sander, *Democratic Disruption in the Age of Social Media: Between Marketized and Structural Conceptions of Human Rights Law*, 32(1) EUR. J. INT'L L. 159 (2021); Myojung Chung and John Wihbey, *Social Media Regulation, Third-Person Effect, and Public Views: A Comparative Study of the United States, the United Kingdom, South Korea, and Mexico*, NEW MEDIA & SOCIETY 146144482211229 (2022).

[79] *See, e.g.,* Controlling the Assault of Non-Solicited Pornography and Marketing Act (CAN-SPAM Act) 2003 (U.S.); Privacy and Electronic Communications (EC Directive) Regulations (PECR) 2003 (UK).

[80] *See, e.g.,* COPPA, *supra* note 25; Online Safety Act, *supra* note 29.

[81] *See, e.g.,* Enhancing Online Safety Act 2015 (Australia).

[82] Gesetz zur Verbesserung der Rechtsdurchsetzung in sozialen Netzwerken (Law to Improve Law Enforcement in Social Networks) (NetzDG) 2017 (Germany).

[83] Jacob Mchangama and Joelle Fiss, The Digital Berlin Wall: How Germany (Accidentally) Created a Prototype for Global Online Censorship (Justitia, Copenhagen, 2019), *at* https://globalfreedomofexpression.columbia.edu/wp-content/uploads/2019/11/Analyse_The-Digital-Berlin-Wall-How-Germany-Accidentally-Created-a-Prototype-for-Global-Online-Censorship.pdf; PAUL CARLS, THE GERMAN STATE'S FIGHT AGAINST HATE SPEECH (2023); Bharath Ganesh, *Content Moderation: Social Media and Countering Online Radicalisation*, *in* THE ROUTLEDGE HANDBOOK ON RADICALISATION AND COUNTERING RADICALISATION 498 (Joel Busher, Leena Malkki, and Sarah Marsden eds., 2024).

[84] Protection from Online Falsehoods and Manipulation Act (POFMA), 2019 (Singapore), ss 10–11.

Though a handful of Covid-19 laws were repealed as the effects of the pandemic waned, the spread of laws addressing misinformation and disinformation continued. Australia, for example, released a draft bill in 2023 on Combatting Misinformation and Disinformation[88] that was hotly debated — including its fair share of fake news.[89] Around the same time, the EU's Digital Services Act came into force,[90] while Britain passed a new Online Safety Act.[91]

All struggle with the problem of how to deal with 'lawful but awful' content online. Australia's bill would have granted its media regulator more power to question platforms on their efforts to combat misinformation. Backlash against perceived threats to free speech led the government to postpone its introduction to Parliament, with promises to 'improve the bill.'[92] The EU legislation avoids defining disinformation, but limits measures on socially harmful (as opposed to 'illegal') content to 'very large online platforms' and 'very large online search engines'[93] — in essence, big tech companies like Google, Meta, and the like. Ofcom, the body tasked with enforcing the new UK law, states that it is 'not responsible for removing online content', but will help ensure that firms have effective systems in place to prevent harm.[94]

---

[85] *See, e.g.,* Corrections Regarding Falsehoods in Mr Leong Mun Wai's Social Media Posts (12 Feb 2024) and Republished by the Online Citizen and Gutzy Asia (Minister for Social and Family Development, Singapore, 15 February 2024), *at* https://www.gov.sg/article/factually150224.

[86] *See, e.g.,* Kirsten Han, *Want to Criticize Singapore? Expect a "Correction Notice"*, N.Y. TIMES, 21 January 2020; Singapore: 'Fake News' Law Curtails Speech (Human Rights Watch, New York, 13 January 2021), *at* https://www.hrw.org/news/2021/01/13/singapore-fake-news-law-curtails-speech; Terence Lee and Walid Jumblatt Abdullah, *(Not) Democratising Through Strength: Core Beliefs and the Institutions of Singapore's People's Action Party*, 28(5) CONTEMPORARY POLITICS 587-610 (2022). Cf Shashi Jayakumar, Benjamin Ang, and Nur Diyanah Anwar, *Fake News and Disinformation: Singapore Perspectives*, *in* DISINFORMATION AND FAKE NEWS 137 (Shashi Jayakumar, Benjamin Ang, and Nur Diyanah Anwar eds., 2021); Alan Chong, *Smart City, Small State: Singapore's Ambitions and Contradictions in Digital Transnational Connectivity*, 74(1) JOURNAL OF INTERNATIONAL AFFAIRS 243 (2021).

[87] *See infra* n 129.

[88] Communications Legislation Amendment (Combatting Misinformation and Disinformation) Bill (Exposure Draft) (Parliament of the Commonwealth of Australia, Canberra, 2023), *at* https://www.infrastructure.gov.au/sites/default/files/documents/communications-legislation-amendment-combatting-misinformation-and-disinformation-bill2023-june2023.pdf.

[89] Amy Remeikis, *Why Is Labor's Bill on Combatting Disinformation so Controversial?*, GUARDIAN, 1 October 2023.

[90] Regulation (EU) 2022/2065 of the European Parliament and of the Council of 19 October 2022 on a Single Market for Digital Services and amending Directive 2000/31/EC (Digital Services Act) 2022 (EU).

[91] Online Safety Act, *supra* note 29.

Such gentle measures may be contrasted with China's more robust approach, where the so-called 'great firewall' is often characterised by over-inclusion.[95] Some years ago, Winnie the Pooh was briefly blocked because of memes comparing him to President Xi Jinping;[96] earlier efforts to limit discussion of the 'Jasmine Revolution' unfolding across the Arab world in 2011 led to a real-world impact on online sales of jasmine tea.[97]

**Alternative Approaches**

Correcting or blocking content is not the only means of addressing the problem. Limiting the *speed* with which false information can be transmitted is another option, analogous to the circuit breakers that protect stock exchanges from high-frequency trading algorithms sending prices spiralling.[98] In India in 2018, WhatsApp began limiting the ability to forward messages after lynch mobs killed several people following rumours circulated on the platform.[99] A study based on data in India, Brazil, and Indonesia showed that such methods can delay the spread of information, but are not effective in blocking the propagation of disinformation campaigns in public groups.[100] Virality disruptors or circuit-breakers have been proposed as a measure to slow down certain classes of information when it threatens to overwhelm a platform.[101]

---

[92] Josh Taylor, *Labor to Overhaul Misinformation Bill after Objections over Freedom of Speech*, GUARDIAN, 13 November 2023.

[93] Digital Services Act, *supra* note 90, arts 33-43.

[94] Online Safety – What Is Ofcom's Role, and What Does It Mean for You? (Ofcom, London, 2023), *at* https://www.ofcom.org.uk/news-centre/2023/online-safety-ofcom-role-and-what-it-means-for-you.

[95] JAMES GRIFFITHS, THE GREAT FIREWALL OF CHINA: HOW TO BUILD AND CONTROL AN ALTERNATIVE VERSION OF THE INTERNET (2019).

[96] *"Oh, bother": Chinese Censors Can't Bear Winnie the Pooh*, STRAITS TIMES, 17 July 2017.

[97] Bruce J. Dickson, *No "Jasmine" for China*, 110(737) CURRENT HISTORY 211 (2011).

[98] *See supra* n 50.

[99] Sankalp Phartiyal and Krishna V. Kurup, *WhatsApp Curbs Message Forwarding in Bid to Deter India Lynch Mobs*, REUTERS, 21 July 2018.

Another platform-based approach is to be more transparent about the provenance of information. One such approach is to require users to verify their identity. China, for example, has required the use of real names of internet users since at least 2012.[102] South Korea briefly had a similar regime, but it was struck down as unconstitutional.[103] Similar systems have occasionally been mooted in other jurisdictions, but few appear to have gained traction outside limited uses — such as accessing specific services or verifying age. In some jurisdictions, such as Germany, the law specifically prohibits forcing the use of real names if this is not necessary for a given service.[104]

More recently, the rise of generative AI and synthetic content has led to debates about whether and how such content itself could be labelled in such a way that consumers would know when they are viewing AI-generated material.[105] Tellingly, the US tech companies that agreed to voluntary watermarking in 2023 limited those commitments to images and video, echoed in the Biden Administration's October 2023 executive order.[106] Synthetic text is nearly impossible to label consistently; as it becomes easier to generate multimedia, it is likely that images and video will go the same way.[107] It has been suggested that such labelling might partially address the well-known problem of 'hallucinations' or confabulations by generative AI systems, potentially pushing responsibility upstream — though the trend at present is merely to include a disclaimer along the lines of 'ChatGPT can make mistakes. Check important info' and leave it at that.

---

[100] Philipe de Freitas Melo et al., "Can WhatsApp Counter Misinformation by Limiting Message Forwarding?" (paper presented at the Complex Networks and Their Applications VIII, Cham, 2020).

[101] *See, e.g.,* Ellen P. Goodman, Digital Information Fidelity and Friction: Crafting a Systems-Level Approach to Transparency (Knight First Amendment Institute, Columbia University, New York, 26 February 2020), *at* https://knightcolumbia.org/content/digital-fidelity-and-friction; Joseph B. Bak-Coleman et al., *Combining Interventions to Reduce the Spread of Viral Misinformation*, 6(10) NATURE HUMAN BEHAVIOUR 1372 (2022/10/01 2022).

[102] King-wa Fu, Chung-hong Chan, and M. Chau, *Assessing Censorship on Microblogs in China: Discriminatory Keyword Analysis and the Real-Name Registration Policy*, 17(3) IEEE INTERNET COMPUTING 42 (2013); Jyh-An Lee and Ching-Yi Liu, *Real-Name Registration Rules and the Fading Digital Anonymity in China*, 25(1) PAC. RIM L. & POL'Y J. 1 (2016).

[103] Keechang Kim, *Korean Internet and "Real Name" Verification Requirement*, 20 KOREA UNIVERSITY LAW REVIEW 87 (2016) (introduced in 2004, struck down in 2012).

[104] Telekommunikation-Telemedien-Datenschutzgesetz (Telecommunications and Telemedia Data Protection Act) (TTDSG) 2021 (Germany), § 19(2); Oliver Noyan, *German Supreme Court Orders Facebook to Allow Pseudonyms*, EURACTIV.DE, 28 January 2022.

[105] *See, e.g.,* Provisions on the Administration of Deep Synthesis of Internet-Based Information Services (互联网信息服务深度合成管理规定) [CLI Code] CLI.4.5145526(EN) 2023 (China); Matt Sheehan, Tracing the Roots of China's AI Regulations (Carnegie Endowment for International Peace, Washington, DC, February 2024), *at* https://carnegieendowment.org/research/2024/02/tracing-the-roots-of-chinas-ai-regulations.

In fact, as synthetic media becomes more common, it may be easier to label content that is human-generated rather than that which is not. Trusted organizations may also watermark images so that users can identify the source of any piece of content.[108] The problem here is that tracking such data may require effort and many users demonstrate little interest in spending the time to verify whether information is true or not. For example, Twitter (prior to its acquisition by Elon Musk) introduced a 'read before you retweet' prompt, which was intended to stop knee-jerk sharing of news based solely on the headline. It appeared to have a positive impact,[109] but was not enough to stop the slide into toxicity.

## 2.3 Consumption?

The ideal, of course, is for users to take responsibility for what they consume and share. Those who grew up watching curated nightly news or scanning a physical newspaper may be mystified by a generation that learns about current events from social media feeds and the next video on TikTok.[110]

---

[106] Executive Order on the Safe, Secure, and Trustworthy Development and Use of Artificial Intelligence (White House, Washington, DC, 30 October 2023).

[107] Sonia Salman, Jawwad Ahmed Shamsi, and Rizwan Qureshi, *Deep Fake Generation and Detection: Issues, Challenges, and Solutions*, 25(1) IT PROFESSIONAL 52 (2023). There are already services such as 'Humanize AI' and 'Undetectable AI' that offer to take AI-produced and render it more 'human-like'. *See* further Jacob Hoffman-Andrew, AI Watermarking Won't Curb Disinformation (Electronic Frontier Foundation (EFF), San Francisco, 5 January 2024), *at* https://www.eff.org/deeplinks/2024/01/ai-watermarking-wont-curb-disinformation.

[108] *See, e.g.,* the work of the Coalition for Content Provenance and Authenticity (C2AP), available at https://c2pa.org

[109] Rebecca Polly Leider, *Do Questions like "Do You Want to Share this Article Without Clicking It First" Help Prevent the Spread of Misinformation?*, NEWS LITERACY MATTERS, 26 April 2023.

[110] Cf JAMES MEESE AND SARA BANNERMAN, THE ALGORITHMIC DISTRIBUTION OF NEWS: POLICY RESPONSES (2022); MICHAEL FILIMOWICZ, INFORMATION DISORDER: ALGORITHMS AND SOCIETY (2023).

Yet concerns about the information diet of the public are as old as democracy itself. Some months before the US Constitution was drafted in 1787, Thomas Jefferson pondered whether it would be better to have a government without newspapers or newspapers without a government. 'I should not hesitate a moment to prefer the latter,' he concluded, making clear that he meant that all citizens should receive those papers and be capable of reading them.[111]

This is not to suggest that mere consumption of material is a suitable subject for legislation. Various countries and organizations are experimenting with different forms of policy interventions, notably educational strategies. Interesting approaches include those that approach consumers' information diets as analogous to nutritional health: poisons are prohibited, alcohol and the like are limited to adults, basic hygiene requirements are intended to protect all, and a labelling regime nudges consumers towards healthful foods and away from sugars and the like.[112] Such strategies will be a vital part of any joined-up effort to limit the impact of misinformation and the like, but go beyond the scope of the present article.[113]

# 3   Emerging Trends in Legislation

The challenge of misinformation, disinformation, and the like demands a multifaceted response. The scale of the problem means that technical interventions — ranging from automated content moderation to exclude the most harmful content from ever being published to mechanisms to flag problematic material — will, on a day-to-day basis, be the most important.[114] Education and other policies will also be a vital part of any strategy to address the impact on individuals and society.[115] But it has also become clear to governments around the world that legislation is also required to manage the flow of content online that is not otherwise covered by existing laws.

---

[111] Thomas Jefferson, *Letter from Thomas Jefferson to Edward Carrington, 16 January 1787*, in 5 THE WORKS OF THOMAS JEFFERSON 251, 253 (Paul Leicester Ford ed., 1904) .

[112] Talia Bulka, *Algorithms and Misinformation: The Constitutional Implications of Regulating Microtargeting*, 32 FORDHAM INTELLECTUAL PROPERTY, MEDIA & ENTERTAINMENT LAW JOURNAL 1107 (2022); Aziz Z. Huq, *International Institutions and Platform-Mediated Misinformation*, 23 CHI. J. INT'L L. 116 (2022).

[113] *See also* Britt et al., *supra* note 56; Marcus Leaning, *An Approach to Digital Literacy Through the Integration of Media and Information Literacy*, 7(2) MEDIA AND COMMUNICATION 4 (2019); THE PSYCHOLOGY OF FAKE NEWS: ACCEPTING, SHARING, AND CORRECTING MISINFORMATION (Rainer Greifeneder, Mariela Jaffe, and Eryn Newman eds., 2020); JOURNALISM RESEARCH THAT MATTERS (Valérie Bélair-Gagnon and Nikki Usher eds., 2021); Kacper T. Gradoń et al., *Countering Misinformation: A Multidisciplinary Approach*, 8(1) BIG DATA & SOC'Y (2021); LESLEY S.J. FARMER, FAKE NEWS IN CONTEXT (2021); David Xiang and Lisa Soleymani Lehmann, *Confronting the Misinformation Pandemic*, 10(3) HEALTH POLICY AND TECHNOLOGY 100520 (2021); MICHAEL HAMELEERS, POPULIST DISINFORMATION IN FRAGMENTED INFORMATION SETTINGS: UNDERSTANDING THE NATURE AND PERSUASIVENESS OF POPULIST AND POST-FACTUAL COMMUNICATION (2022).

To study emerging trends in legislation, we searched for all relevant laws introduced as of 31 December 2023.[116] The scope was defined as laws addressing online misinformation, disinformation, and mal-information however termed (including fake news, falsehoods, hostile information campaigns, and analogous terms). General laws governing obscenity, incitement, harassment, defamation, and national security were not included in favour of a focus on digital or online activity. Several existing databases provided a starting point,[117] but the search systematically examined all 193 member states of the United Nations by using keyword searches for legislation based on the above terms.[118]

The focus was on national legislation, meaning that it is possible that some state or provincial laws are omitted. Regional organizations, most prominently the European Union, are not included in this analysis, though the adoption of the Digital Services Act and Digital

---

[114] There are few publicly available estimates of the amount of content that is automatically blocked by social media companies. Meta provides aggregated percentages of content removed in its Community Standards Enforcement Report (*see* https://transparency.meta.com/policies/improving/proactive-rate-metric); Google/YouTube reports on the number of videos and channels removed (*see* https://transparencyreport.google.com/youtube-policy/removals).

[115] *See supra* n 113.

[116] Data collection was conducted together with three law student research assistants. Existing compilations were reviewed, with each of the three students then taking a region of the world for further study by individual country. A final Google search was conducted using the strings (a) "[country name]" "disinformation" "law"; (b) "[country name]" "misinformation" "law"; and (c) "[country name]" "fake news" "law". Results were tabulated in a shared spreadsheet with spot validation by the author.

[117] Jacob N. Shapiro, Jan Oledan, and Samikshya Siwakoti, ESOC COVID-19 Misinformation Dataset (Empirical Studies of Conflict, Princeton, 2020), *at* https://esoc.princeton.edu/publications/esoc-covid-19-misinformation-dataset; Gabrielle Lim and Samantha Bradshaw, Chilling Legislation: Tracking the Impact of "Fake News" Laws on Press Freedom Internationally (Center for International Media Assistance, Washington, DC, 2023), *at* https://www.cima.ned.org/publication/chilling-legislation/; Daniel Funke and Daniela Flamini, A Guide to Anti-Misinformation Actions Around the World (Poynter, St Petersburg, FL, 2023), *at* https://www.poynter.org/ifcn/anti-misinformation-actions/.

[118] Some data was included on jurisdictions such as Taiwan, as well as regional and intergovernmental approaches that may be the subject of future analysis.

Markets Act will make that more important in future studies.[119] It is also possible that best efforts to include languages other than English were unsuccessful in gathering a complete dataset, and that reliance on online materials does not fully represent legislation in states lacking elaborate digital infrastructure. A further qualification is that several laws prior to 1995 were included in the initial dataset, given their ongoing application to online as well as offline speech — such as legislation covering broadcasting as well as several criminal codes. To focus on legislation either specifically targeting online activity, or adopted in the knowledge that online activity was an increasingly important forum, the data presented here takes 1995 as the starting point.[120] The full dataset is published as an appendix to this article and can doubtless be further improved, as well as kept up to date.

Despite all these limitations, several interesting findings emerge from the data and are presented here in three broad areas. The first is the growing number of such laws and their spread around the world. Secondly, the coverage of those laws has also evolved in terms of the perceived threats and the actors targeted. A third aspect of the turn to legislation is the expanding powers given to regulatory authorities to enforce the new laws.

## 3.1 Growth

The first and most obvious finding is that the number of such laws has grown enormously in recent years. As Figure 1 shows, the absolute number of statutes adopted worldwide more than tripled from 45 in 2016 to 151 in 2023.[121] As indicated in the introduction, this corresponds to the period in which misinformation and disinformation became widely discussed topics, particularly in the context of the election of Donald Trump as US President and the Brexit vote in 2016, the Covid-19 pandemic of 2020, and the rise of synthetic media including generative AI.[122]

---

[119] Digital Services Act, *supra* note 90; Regulation (EU) 2022/1925 of the European Parliament and of the Council of 14 September 2022 on Contestable and Fair Markets in the Digital Sector and Amending Directives (EU) 2019/1937 and (EU) 2020/1828 (Digital Markets Act) 2022 (EU).

[120] Though only 0.04% of the world's population had access to the internet in 1995, that was the year in which restrictions on commercial traffic were lifted and usage became more widespread. *See generally* MICHAEL A. BANKS, ON THE WAY TO THE WEB: THE SECRET HISTORY OF THE INTERNET AND ITS FOUNDERS (2008). In addition, laws collected in the dataset from 1990 to 1994 included only amendments to the criminal codes of Nigeria (1990), Mozambique (1991), and Peru (1991) that addressed false information but made no reference to digital or online communications.

[121] Note that these figures include some countries that adopted more than one law.

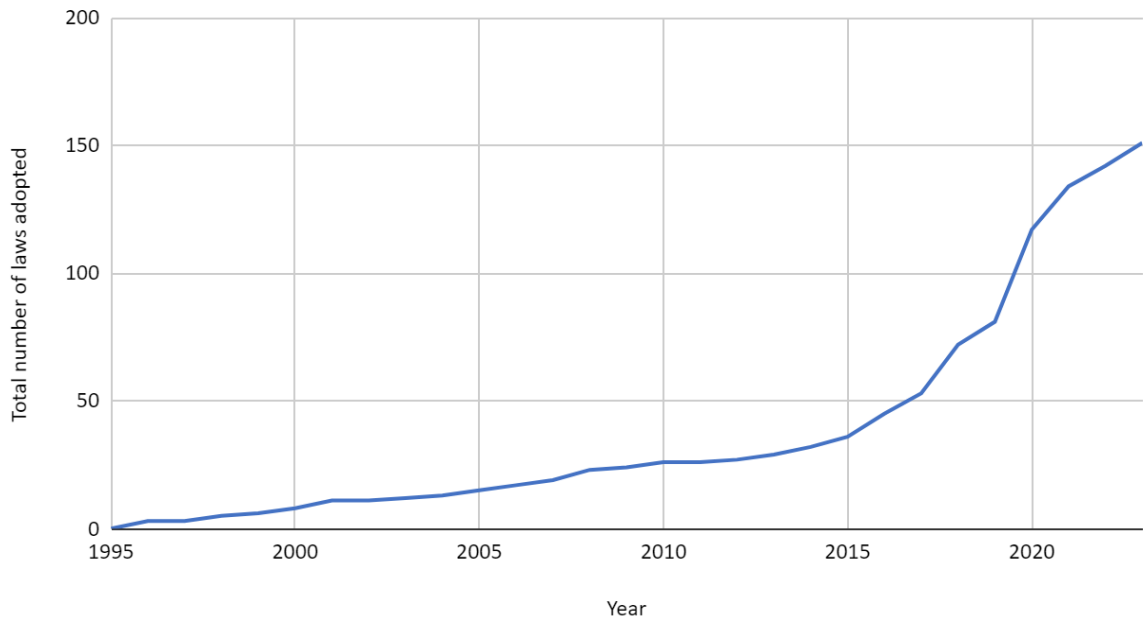

Figure 1: Increase in total number of national laws passed since 1995

Geographically, much of the early growth was in Africa and Asia. Figure 2 breaks growth down by the percentage of states in each of the five UN regional groupings[123] that had adopted at least one law by a given year. Recent growth has, however, been steepest in the Western Europe and Others group: where only two countries having applicable laws in 2016 (the United States and Australia), that figure rose to 14 in 2023. It is not claimed here that regional clustering or convergence is established by these trends, though there is some evidence that this has happened in other areas of law, notably in the context of the European Union and its neighbouring states.[124]

---

[122] *See supra* nn 7–10.

[123] United Nations, Regional groups of Member States, https://www.un.org/dgacm/en/content/regional-groups

[124] Meryll Dean, *Bridging the Gap: Humanitarian Protection and the Convergence of Laws in Europe*, 20(1) EUROPEAN LAW JOURNAL 34 (2014). Cf Lilian Chenwi, *The Right to Adequate Housing in the African Regional Human Rights System: Convergence or Divergence Between the African Commission and South African Approaches*, 17(si-1) LAW, DEMOCRACY & DEVELOPMENT 342 (2013); Mary Crock, *Shadow Plays, Shifting Sands and International Refugee Law: Convergences in the Asia-Pacific*, 63(2) INT'L & COMP. L.Q. 247 (2014).

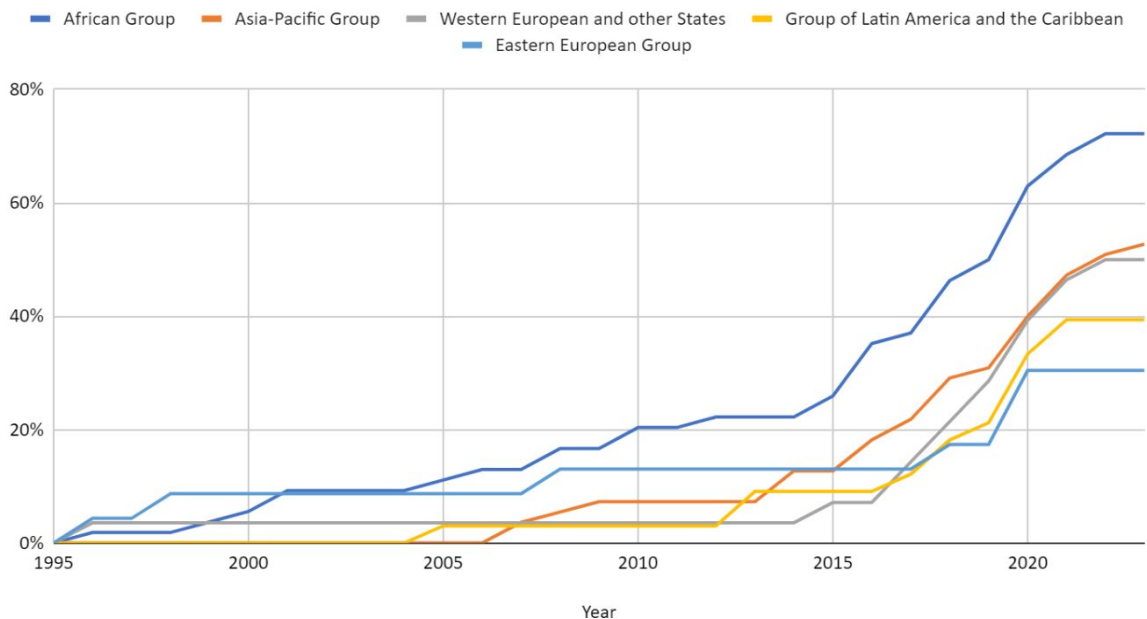

*Figure 2: Percentage of states within UN regional groupings that had adopted laws since 1995*

Interestingly, the adoption of laws does not appear to have been driven by internet penetration. Figure 3 plots the year in which a country's first law on the topic was adopted against internet penetration as a percentage of the population in 2016.[125] Indeed, the slight upward slope on the trend line suggests a modest correlation between *lower* internet penetration and early adoption. Many factors shape the decisions to legislate, including the balance between moderating the risks of online content as against concerns about limiting innovation or driving it elsewhere.[126]

---

[125] Individuals using the Internet (% of population) (World Bank, Washington, DC, 2022), *at* https://data.worldbank.org/indicator/IT.NET.USER.ZS. The year 2016 was chosen as it has more complete data than subsequent years.

[126] Chris Marsden, Trisha Meyer, and Ian Brown, *Platform Values and Democratic Elections: How Can the Law Regulate Digital Disinformation?*, 36 COMPUTER L. & SECURITY REP. 105373 (2020). Cf SIMON CHESTERMAN, WE, THE ROBOTS? REGULATING ARTIFICIAL INTELLIGENCE AND THE LIMITS OF THE LAW 180-85 (2021).

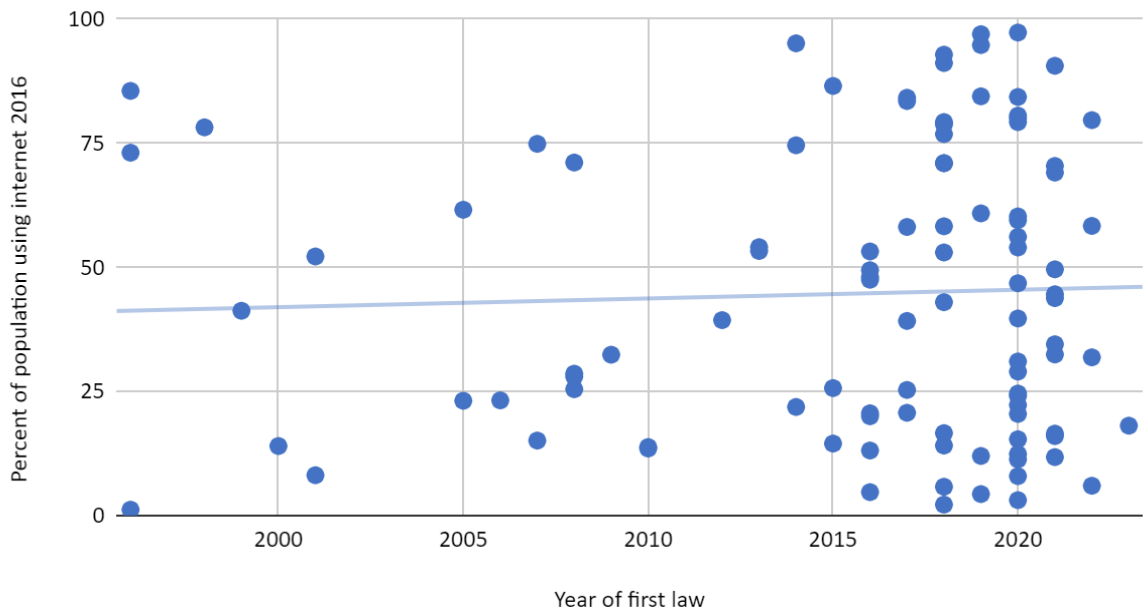

*Figure 3: Year of first law passed, contrasted with internet penetration*

Nor does the level of economic development appear to have been a major factor in adopting such laws. On the contrary, as shown in Figure 4, laws tended to be adopted first in countries with lower gross domestic product per capita, though the low gradient on the line suggests that the correlation is at best a weak one.[127]

---

[127] GDP per capita (current US$) (World Bank, Washington, DC, 2022), *at* https://data.worldbank.org/indicator/NY.GDP.PCAP.CD. As for internet penetration, data from 2016 was chosen for completeness. Data for Eritrea added from Statista as no data was available for 2016 in the World Bank database. https://www.statista.com/statistics/510486/gross-domestic-product-gdp-per-capita-in-eritrea/

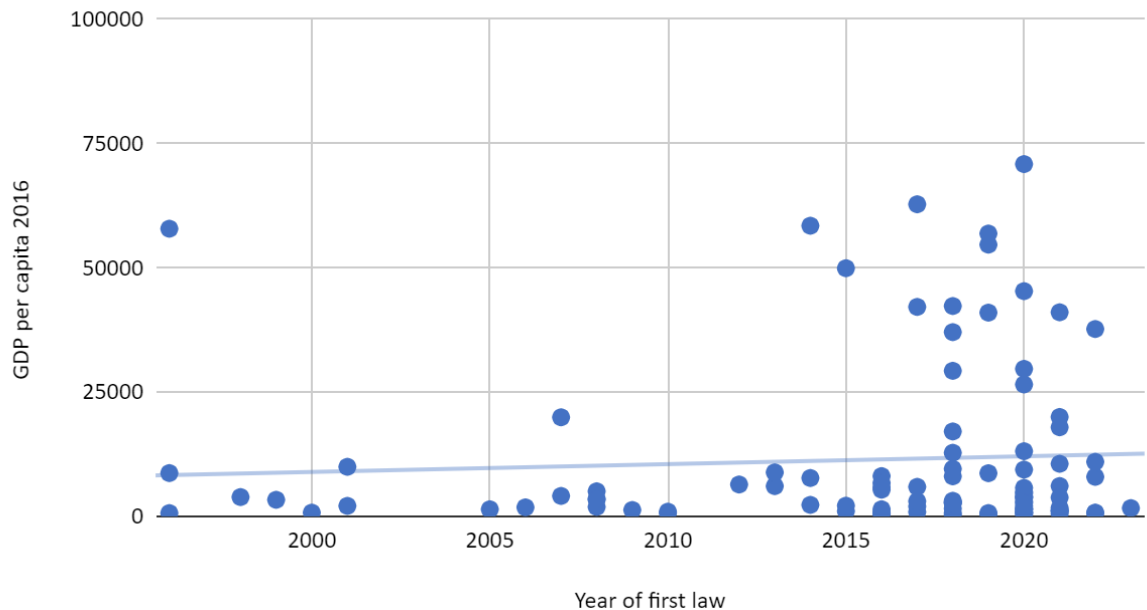

*Figure 4: Year of first law passed, contrasted with GDP per capita*

As in the case of internet penetration, it is important to note that this data merely indicates that a law was passed. The United States, for example, is the outlier as a high-GDP per capita country that adopted an early law — though, as discussed earlier, that 1996 law was primarily intended to *limit* the responsibility of internet platforms for the content posted on them.[128]

## 3.2 Coverage

Not all laws specify a clear purpose. Amendments to criminal codes, for example, do not always disclose the public policy behind criminalizing certain activities. To the extent that such purposes are disclosed, however, Figure 5 shows that the most commonly articulated in the legislation adopted since 1995 is national security and/or counterterrorism. It is also striking — if unsurprising — that the number of laws focused on public health tripled from 11 in 2019 to 33 in 2020, almost all explicitly linked to the Covid-19 pandemic of that year. (At least eight of these laws were later repealed.[129]) Note that several pieces of legislation mention more than one term, which would each be counted for the purposes of the chart.

---

[128] *See supra* n 69.

[129] Bolivia (two decrees), Bosnia and Herzegovina, Hungary, the Philippines, and Romania repealed their laws in 2020; Thailand in 2022; Jordan in 2023. Taiwan is not included in the 'country' dataset, but adopted the Special

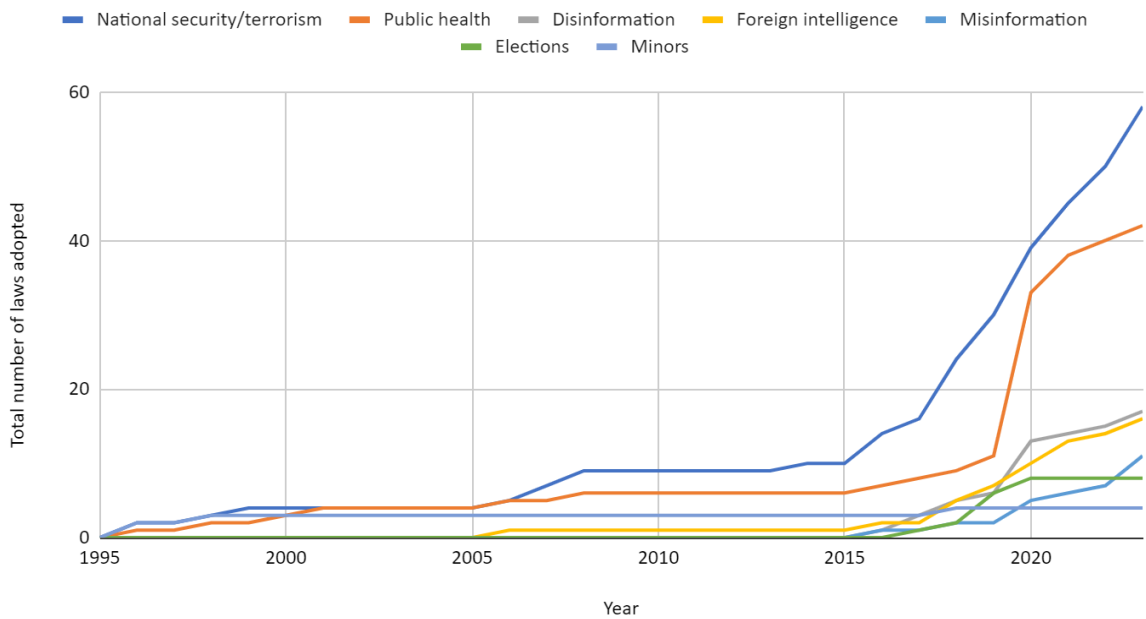

*Figure 5: Coverage of selected terms in laws adopted*

The targets of laws are not always specified either, though they typically focus on the creators of inappropriate content and knowing sharers, as shown in Figure 6. Media organizations are also included in some laws, though they may also be subject to broader regulations concerning broadcasters. A focus on platforms is of growing interest, with the notable example of the EU's legislation discussed earlier.[130] Innocent sharers of false content are the least likely to be within scope, often with more limited sanctions.[131]

---

Act for Prevention, Relief and Revitalization Measures for Severe Pneumonia with Novel Pathogens in 2020 and repealed it in 2023.

[130] *See supra* n 90.

[131] *See also* Shalini Talwar et al., *Why Do People Share Fake News? Associations Between the Dark Side of Social Media Use and Fake News Sharing Behavior*, 51 JOURNAL OF RETAILING AND CONSUMER SERVICES 72 (2019)

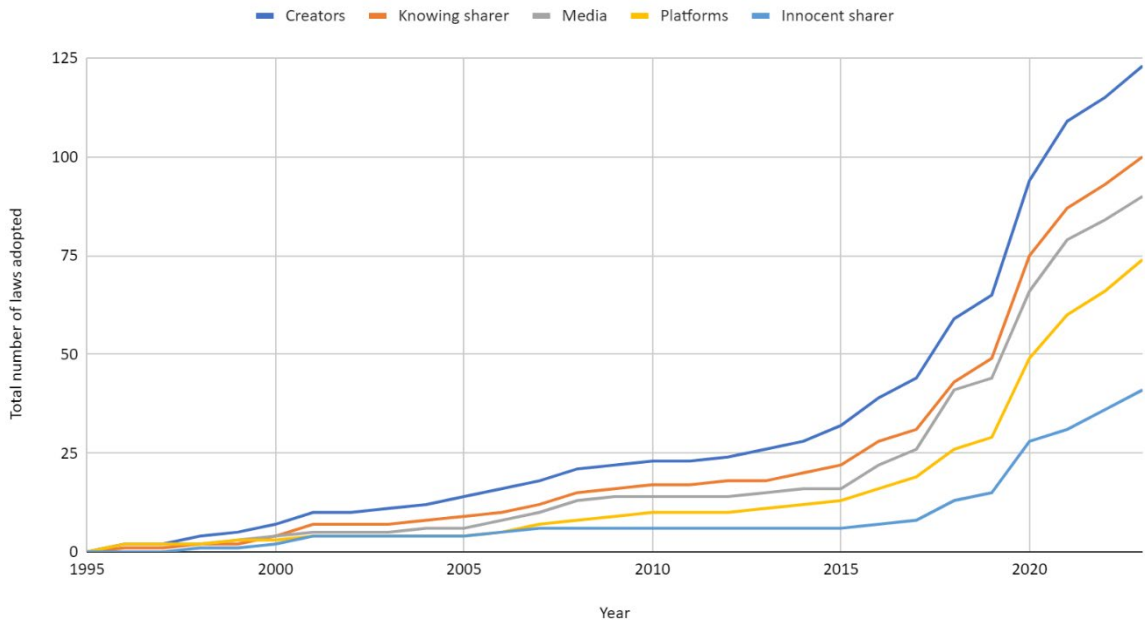

*Figure 6: Number of laws targeting specified actors*

## 3.3 Powers

The present study does not purport to analyse the effectiveness or impact of the laws that are being adopted.[132] Nonetheless, the scope of application and the powers given to regulators are proxies for the willingness to devote resources to the problem and to impose limits on unfettered free speech online.[133] Note, of course, that expansive or limited regulatory powers are affected by domestic constitutional and political factors. It is therefore interesting to see whether laws tended to be adopted earlier or later in countries that are less 'free'. One such measure is Freedom House's global freedom Index.[134] The index has been the subject of criticism,[135] but broadly reflects political rights and civil liberties. The bottom third of the ranking is classed as 'not free', the top third as 'free', and the middle third as 'partly free'. Figure 7 plots the year in which a law was first introduced in a given country against its score out of 100 in the Freedom Index for 2024. Though no causal connection is asserted here, the tendency of states with fewer protections of free speech to be more willing to police speech online is not surprising.[136] What may be worth studying further is the speed with which otherwise 'free' countries followed suit in legislating, particularly after 2016.

---

[132] *See supra* n 12.

[133] Studies in the securities law sphere have used the formal powers of regulators as proxies. *See, e.g.,* Rafael La Porta, Florencio Lopez-De-Silanes, and Andrei Shleifer, *What Works in Securities Laws?*, 61(1) J. Fin. 1 (2006); Simeon Djankov et al., *The Law and Economics of Self-Dealing*, 88(3) J. Fin. Econ. 430 (2008). For a critique of this approach, *see* Howell E. Jackson and Mark J. Roe, *Public and Private Enforcement of Securities Laws: Resource-based Evidence*, 93(2) J. Fin. Econ. 207-38 (2009) (arguing that resources, such as staffing levels, offer a better measure of regulatory intensity).

[134] Freedom in the World (Freedom House, Washington, DC, 2024), *at* https://freedomhouse.org/report/freedom-world#Data.

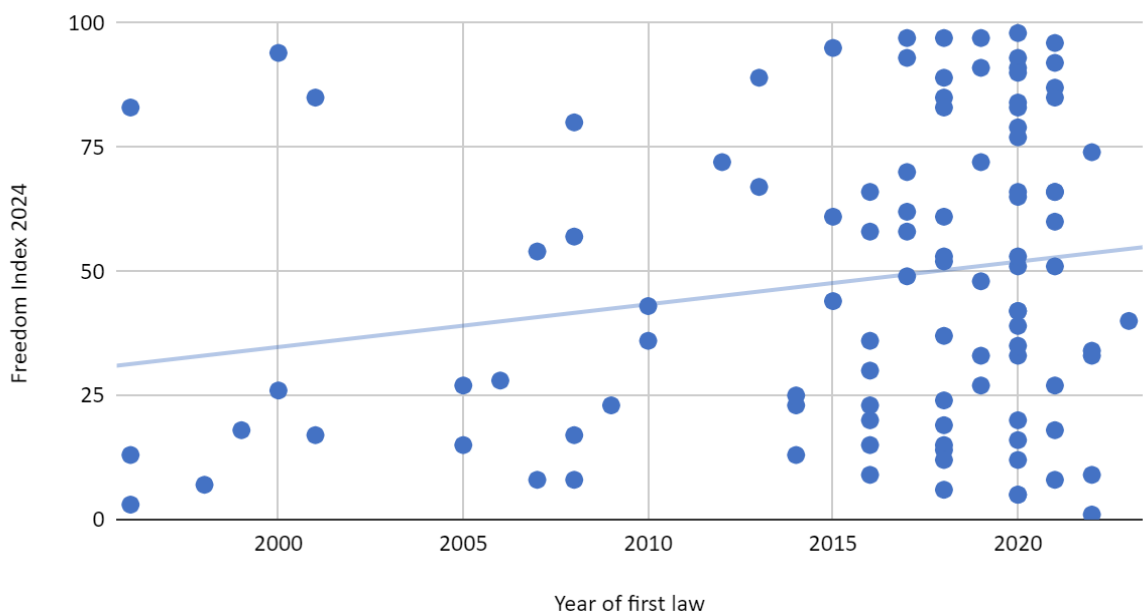

*Figure 7: Year of first law adopted, contrasted with Freedom Index ranking*

This is consistent with Figure 8, which shows the percentage of not free, partly free, and free countries that have adopted laws over time. Though 'not free' countries remain most likely to have adopted such laws, 'free' countries are closing that gap. Since the percentage of countries must inevitably taper, that is to be expected, though once again the uptick in legislative activity since 2016 is significant.

---

[135] *See, e.g.,* Diego Giannone, *Political and Ideological Aspects in the Measurement of Democracy: The Freedom House Case*, 17(1) DEMOCRATIZATION 68 (2010); Nils D. Steiner, *Comparing Freedom House Democracy Scores to Alternative Indices and Testing for Political Bias: Are US Allies Rated as More Democratic by Freedom House?*, 18(4) JOURNAL OF COMPARATIVE POLICY ANALYSIS 329 (2016).

[136] Lee and Abdullah, *supra* note 86.

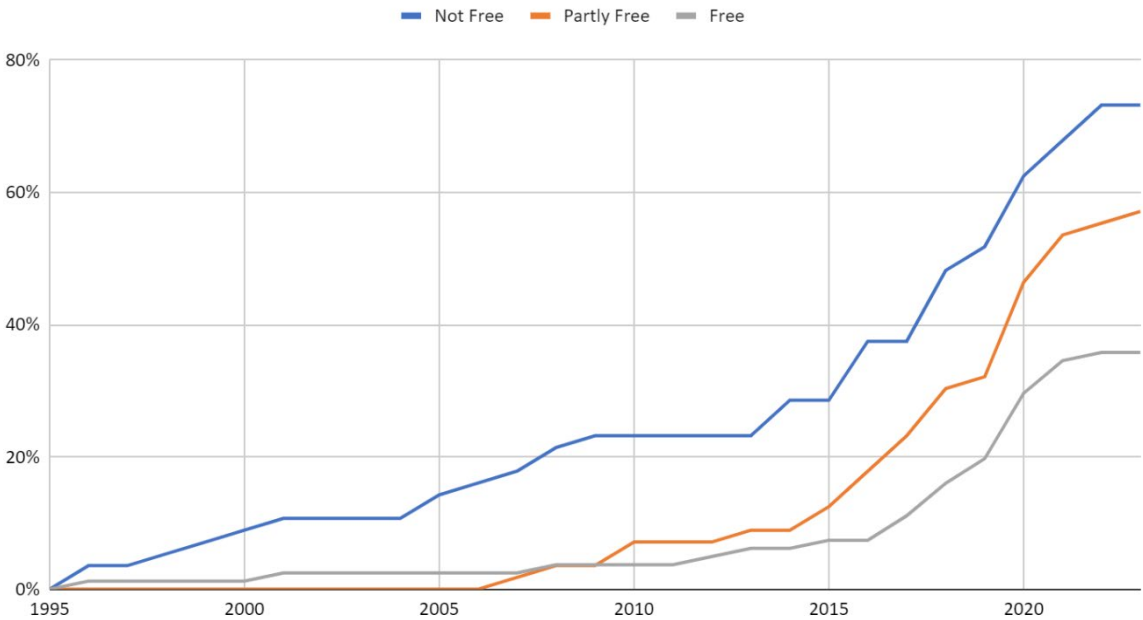

*Figure 8: Percentage of free, partly free, and not free states that have adopted laws*

It bears repeating that the existence of a law is not a measure of its impact. The various laws grant different forms of power to the relevant regulatory authorities. Figure 9 shows various powers included in the legislation captured by the survey, with fines and prison terms being the most common method of enforcing criminal law generally. Though some laws (such as Singapore's POFMA, discussed earlier[137]) provide for determinations of truth or falsity and the publication of corrections, removal of undesirable content is the more common remedy.

---

[137] *See supra* n 84.

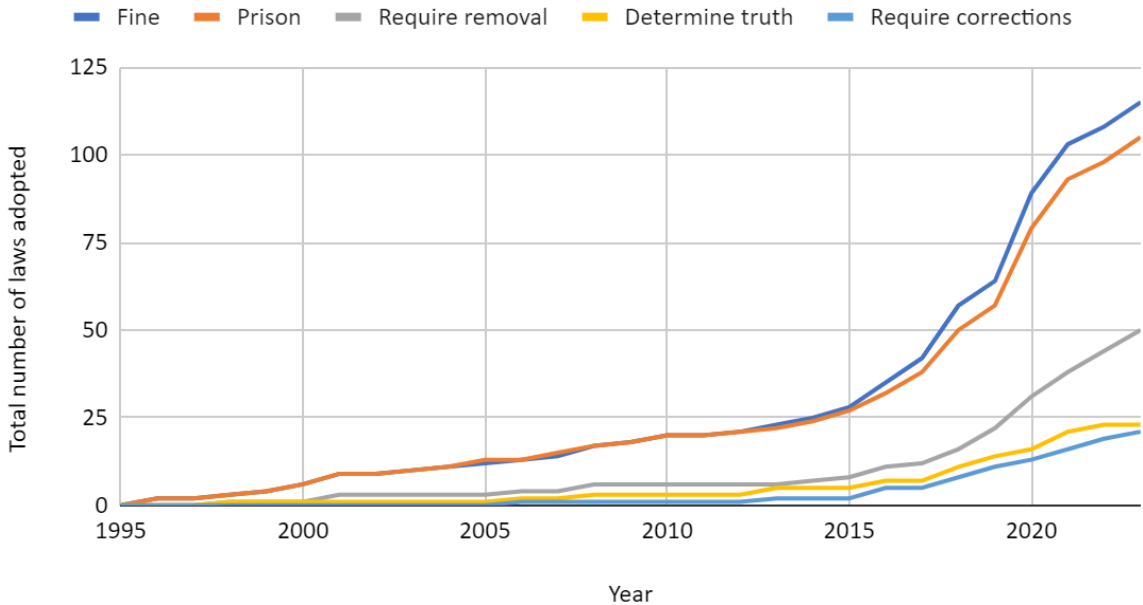

*Figure 9: Number of laws granting specified powers*

To compare the relative powers granted by laws across the many countries surveyed, a composite power measure was created based on the breadth of coverage and the enforcement options available. The composite power figure comprises two parts. First, in terms of coverage, (a) one point was added for covering each of creators, platforms, and media; (b) two points were added for coverage of knowing sharers; (c) five points were added for coverage of innocent sharers. Secondly, in terms of powers, (d) one point was added for powers including corrections; (e) two points for powers of removal; (f) three points for powers to fine; (g) four points for powers to imprison.

This created a numerical measure with a maximum of 20 that offers a rough comparison of the powers arrogated to government.[138] As Figure 10 shows, countries classed as 'less free' under the Freedom House measure (that is, with a lower Freedom Index score) are more likely to have adopted laws with wider scope and greater powers of enforcement.

---

[138] The maximum score of 20 was reached by China, Kyrgyzstan, Singapore, Syria, Thailand, and the United Arab Emirates.

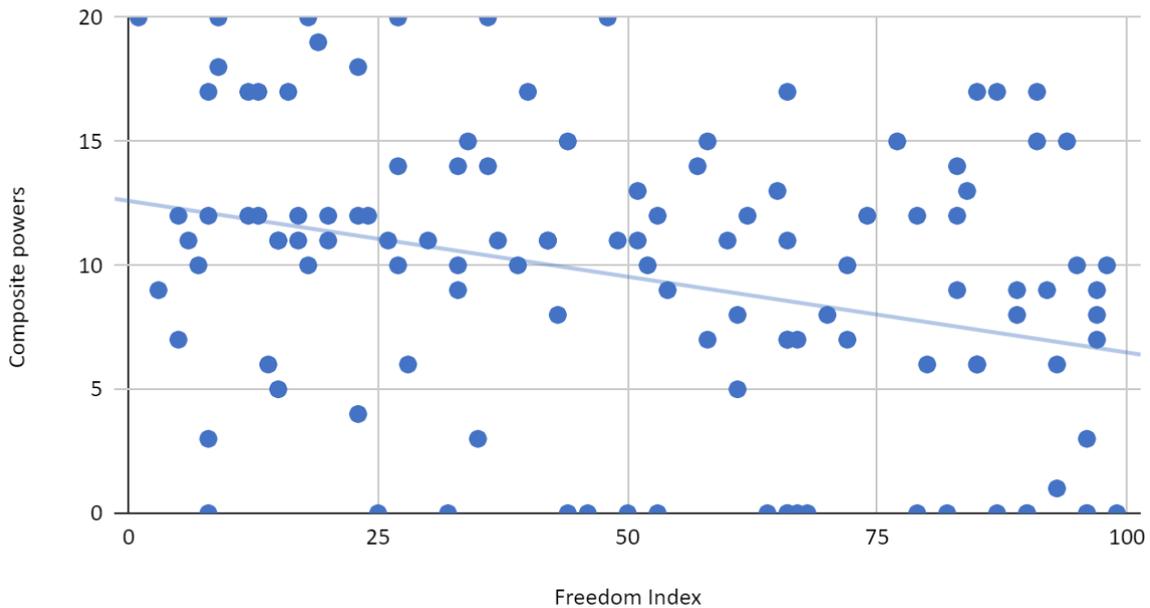

*Figure 10: Composite powers granted by law in a given state, contrasted with Freedom Index rating*

An alternative measure was to average out the powers granted to governments in each of the UN regional groupings. As Figure 11 shows, the average composite power is increasing — that is, independently of the number of laws, the power of those laws is growing — with Western Europe and Others having the lowest average power, while Eastern European and, increasingly, Asian states have given greater sway to their regulators.[139]

---

[139] The Eastern European figure is skewed somewhat by small numbers as, until 2008, laws had been adopted only by Russia (1996, 2001, 2006) and Azerbaijan (1998).

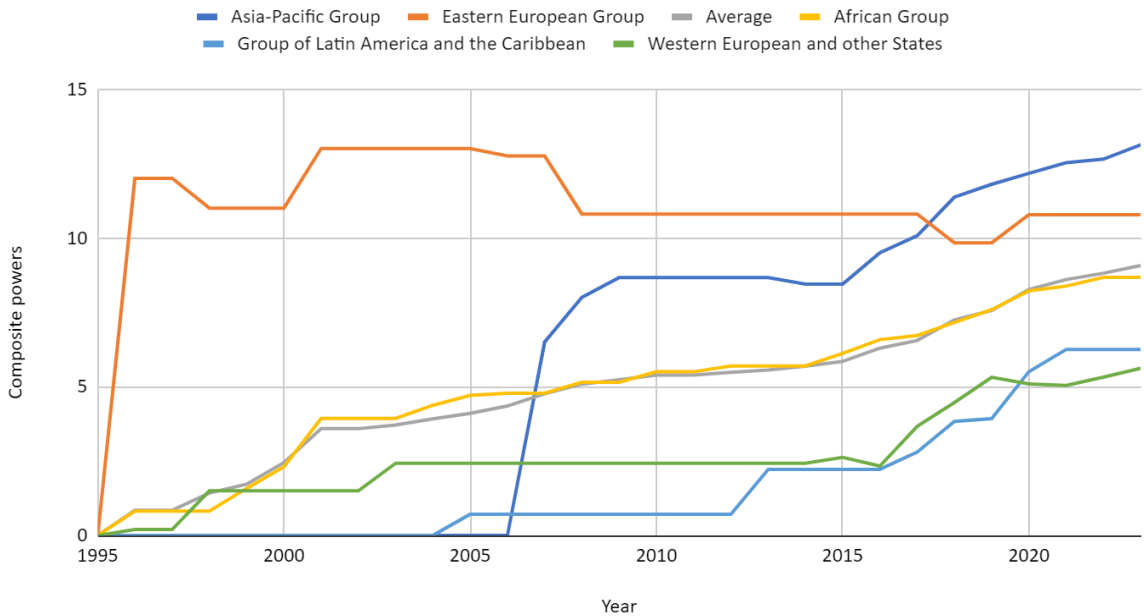

*Figure 11: Composite powers granted by law, over time and by region*

There are obvious limitations to a quantitative study such as this. Some statutes may have been missed or might be categorized differently. Yet the overall trend patterns reflect growing awareness of the challenges posed by misinformation, disinformation, and mal-information and a willingness to adopt legislation to confront it. Though early adopters tended to be African and Asian countries, poorer countries, and less 'free' countries, the phenomenon is now global as the question moves from *whether* to regulate online content to *how* it should be regulated.

# 4 Conclusion

The libertarian streak among technology entrepreneurs runs deep. Bill Gates is said to have bragged in the early years of Microsoft that the company did not even have an office in Washington, DC — he wanted nothing from the government except to be left alone.[140] This was representative of the wider culture in Silicon Valley: most saw their work as undeserving of regulation; a good many deemed themselves morally superior to the governments that might presume to impose it.[141]

---

[140] Nat Levy, *Bill Gates says Microsoft — not Google — Should Have Built the Dominant Mobile Operating System*, GEEKWIRE, 24 June 2019; Gates later came to regret the statement, which might have been seen as taunting regulators.

Culture, in the end, caught up with them. 'Post-truth' was chosen as word of the year by Oxford Dictionaries in 2016.[142] Twelve months later, Collins gave that honour to 'fake news'.[143] 'Misinformation' was anointed in 2018 by both the *Washington Post* and Dictionary.com.[144] The view that the digital commons could be left to regulate itself was increasingly untenable. The Covid-19 pandemic — spawning neologisms such as 'infodemic'[145] — confirmed a turning point in the willingness of regulators to step in.

Though governments around the world had long been aware of the risks associated with online communications, those who legislated early tended to be those with fewer constraints on government power generally. If anything, there appears to have been a negative corelation with internet access in those countries. That is now changing. Legislation is becoming more common and more robust.

This article sought to identify the kinds of content (lawful but awful) that might be appropriate targets for legislation and the possible responses. It then introduced a dataset of laws that indicate some early trends in legislative action. As the pace of change in this area increases — in particular, with the prospect of ever-greater quantities of ever-more realistic synthetic material online — the need for a joined-up strategy is clear. This will need to cover the information lifecycle, from production and distribution to consumption. Industry standards and other forms of supervised self-regulation will be key at the level of platform and technology; educational and social policies will be vital at the level of users. But state

---

[141] *See, e.g.,* Emanuel Moss and Jacob Metcalf, *The Ethical Dilemma at the Heart of Big Tech Companies*, Harvard Business Review, 14 November 2019. Cf David Broockman, Greg F. Ferenstein, and Neil Malhotra, *Predispositions and the Political Behavior of American Economic Elites: Evidence from Technology Entrepreneurs*, 63 Am. J. Pol. Sci. 212 (2019).

[142] Alison Flood, *"Post-Truth" Named Word of the Year by Oxford Dictionaries*, Guardian, 15 November 2016. On the history of 'post-truth', *see* Gerald Posselt and Sergej Seitz, *Truth and Its Political Forms: An Explorative Cartography*, forthcoming Contemporary Political Theory (2023).

[143] Alison Flood, *Fake News Is "Very Real" Word of the Year for 2017*, Guardian, 2 November 2017.

[144] Valerie Strauss, *Word of the Year: Misinformation. Here's Why*, Wash. Post, 10 December 2018; *"Misinformation" Picked as Word of the Year by Dictionary.com*, Guardian, 26 November 2018.

[145] *The COVID-19 Infodemic*, 20(8) Lancet Infectious Diseases 875 (2020).

action will also be essential, including the ability to wield the 'regulatory hammer', when necessary.[146]

A growing number of states have clearly embraced this notion in theory, though more research is needed to see whether it can and will make a difference to the proliferation of misinformation, disinformation, and mal-information in practice.

# 5 Appendix

If appropriate, the complete dataset can be included as an appendix to the article.

---

[146] Margot E. Kaminski, *Binary Governance: Lessons from the GDPR's Approach to Algorithmic Accountability*, 92 SOUTHERN CALIFORNIA LAW REVIEW 1529, 1564 (2019).